\documentclass[twocolumn,prd,amsmath,amssymb]{revtex4}
\usepackage{graphicx}

\renewcommand{\vec}[1]{\boldsymbol{\mathrm{#1}}}

\begin{document}

\title{The masses and shadows of the black holes Sagittarius A* and M87* in modified gravity (MOG)}

\author{J. W. Moffat$^{\star}$ and
V. T. Toth$^\star$\\~\\
{\rm
\footnotesize
$^\star$Perimeter Institute for Theoretical Physics, Waterloo, Ontario N2L 2Y5, Canada}}

\begin{abstract}
We present calculations for the anticipated shadow sizes of Sgr A* and the supermassive black hole in the galaxy M87 in the context of the MOG modified theory of gravity (also known as Scalar--Tensor--Vector--Gravity, or STVG). We demonstrate that mass estimates derived from stellar and gas dynamics in the vicinity of these black holes are the Newtonian masses of the black holes even in the MOG theory. Consequently, shadow sizes increase as a function of the key dimensionless MOG parameter $\alpha$ that characterizes the variable gravitational coupling coefficient $G$, and may offer an observational means to distinguish the MOG theory from standard general relativity.
\end{abstract}


\maketitle

\section{Introduction}

Following up on previous investigations \cite{Moffat2015a,Moffat2015b,Lee2017,Guo2018,Sheoran2018,Wang2019}, we examine the angular size of the shadow cast by the photon spheres of supermassive black holes such as Sagittarius A* (Sgr A*) in the central region of the Milky Way and the black hole in the galaxy M87, in the context of the MOG modified theory of gravity \cite{Moffat2006a}, and the dependence of this apparent size on the theory's parameters.

The theory combines a tensorial gravitational field $g_{\mu\nu}$ and a variable gravitational coefficient $G=G_N(1+\alpha)$ with a massive vector field $\phi_\mu$ associated with a gravitational source charge $Q_g=\sqrt{\alpha G_N}M$ that is proportional to mass and yields a repulsive force, which cancels out excess gravity at short range ($<\cal{O}({\rm kpc})$), making the short-range behavior Newtonian. The test particle equation of motion, derived from the field equations of the theory, in the gravitational field of a compact mass $M$, is given by \cite{Moffat2006a}:
\begin{align}
\ddot{\vec{r}}=-\frac{G_NM}{r^3}\left[1+\alpha-\alpha e^{-\mu r}(1+\mu r)\right]\vec{r},
\end{align}
where $\alpha={\cal O}(1)$ and $\mu={\cal O}({\rm kpc}^{-1})$ are parameters related to the theory's scalar fields, and $G_N$ is Newton's constant of gravitation. This equation of motion remains valid everywhere so long as $v^2/c^2\ll 1$ and $GM/c^2r\ll 1$. For compact sources of mass $M$, a semi-empirical derivation \cite{Moffat2007e} yields the approximate formula $\alpha\simeq\alpha_\infty M/(\sqrt{M}+E)^2$ with $\alpha_\infty={\cal O}(10)$ and $E\sim 25000~M_\odot^{1/2}$, which seems to agree well with galaxy rotation data \cite{GreenMoffat2019}, but we do not know if this weak field approximation is valid for an object as compact as a black hole.

\section{MOG and the photon sphere}

The source mass $M$ that appears in this modified Newtonian acceleration law is the gravitating object's ``Newtonian'' mass. This is to be distinguished from the so-called Arnowitt--Deser--Misner (ADM) mass, $M_{\rm ADM}=(1+\alpha)M$ \cite{Sheoran2018}.

In the case when $r\ll\mu^{-1}$, the MOG equation of motion reduces to the Newtonian form:
\begin{align}
\ddot{\vec{r}}=-\frac{G_NM}{r^3}\vec{r}.\label{eq:shortrange}
\end{align}

The conditions of validity for Eq.~(\ref{eq:shortrange}) are manifestly satisfied for stars in tight orbit around the Milky Way's supermassive black hole, Sgr A*. The closest known star to Sgr A*, S2 (aka. S0-2), reached an estimated speed of 7650 km/s when it approached Sgr A* within 120 astronomical units (AU) \cite{GravColl2018,Doeaav8137}. Even during this close approach, the Newtonian and MOG equations of motion differ only by much less than one part in a thousand and, therefore, the orbits are accurately characterized using the Newtonian equation of motion (\ref{eq:shortrange}).

The immediate consequence of this is that the mass of Sgr A*, as determined from these stellar orbits, is the same Newtonian mass that one would use in Newtonian gravity or general relativity.

Given that, despite the accuracy to which these orbits are known, the Keplerian orbits of stars around Sgr A* are not sensitive to differences between the Newtonian and MOG equations of motion (although we note that, in the future, accurate measurements of this star's periastron shift may help distinguish between general relativity and alternative theories of gravity), the question arises: Are there observable properties of supermassive black holes that show such sensitivity and may help confine the theory's parameters? Mergers of supermassive black holes are a natural candidate to constrain the value of the $\alpha$ parameter \cite{TianQin2019MOG}. However, no such merger has taken place, nor do we have the ability to observe such a merger in the foreseeable future. The other possibility is the apparent size of the supermassive black hole's photon sphere, which determines its ``shadow''. We can ask how, in the context of the MOG theory, the shadow changes as a function of the MOG parameter $\alpha$.


To answer this question, we turn to the Kerr--MOG metric, which is a variation of the Kerr--Newman metric with the substitution $G_N\to(1+\alpha)G_N=G$. Inferred from the MOG gravitational field equations, this metric has the form in Boyer--Lindquist coordinates $r$, $\theta$, $\phi$ \cite{Moffat2015a,Moffat2015b}:
\begin{widetext}
\begin{align}
ds^2=\left(1-\frac{r_sr-r_Q^2}{\rho^2}\right)dt^2-\left[r^2+a^2+a^2\sin^2\theta\left(\frac{r_sr-r_Q^2}{\rho^2}\right)\right]\sin^2\theta d\phi^2+2\sin^2\theta\left(\frac{r_sr-r_Q^2}{\rho^2}\right)dtd\phi-\frac{\rho^2}{\Delta}dr^2-\rho^2d\theta^2,
\end{align}
where $\rho^2=r^2+a^2\cos^2\theta$, $\Delta=r^2-r_sr+a^2+r_Q^2$, $a$ is the angular momentum per unit mass, the Schwarzschild--MOG radius is given by $r_s=2(1+\alpha)G_NM$ and the length scale associated with the vector charge is $r_Q=\sqrt{\alpha(1+\alpha)}G_NM$. As in the case of the Kerr--Newman metric, rotation is extremal when $a\to GM=(1+\alpha)G_NM$ \cite{Guo2018}.

While recognizing that rotation ($a>0$) distorts the shape of the photon sphere and the black hole shadow, these distortions remain small and likely unobservable even for near extremal rotation. (This is not a feature unique to MOG; the behavior is identical in the Kerr--Newman case in general relativity.) Therefore, for simplicity, we consider only the non-rotating case.


\begin{figure}
\includegraphics{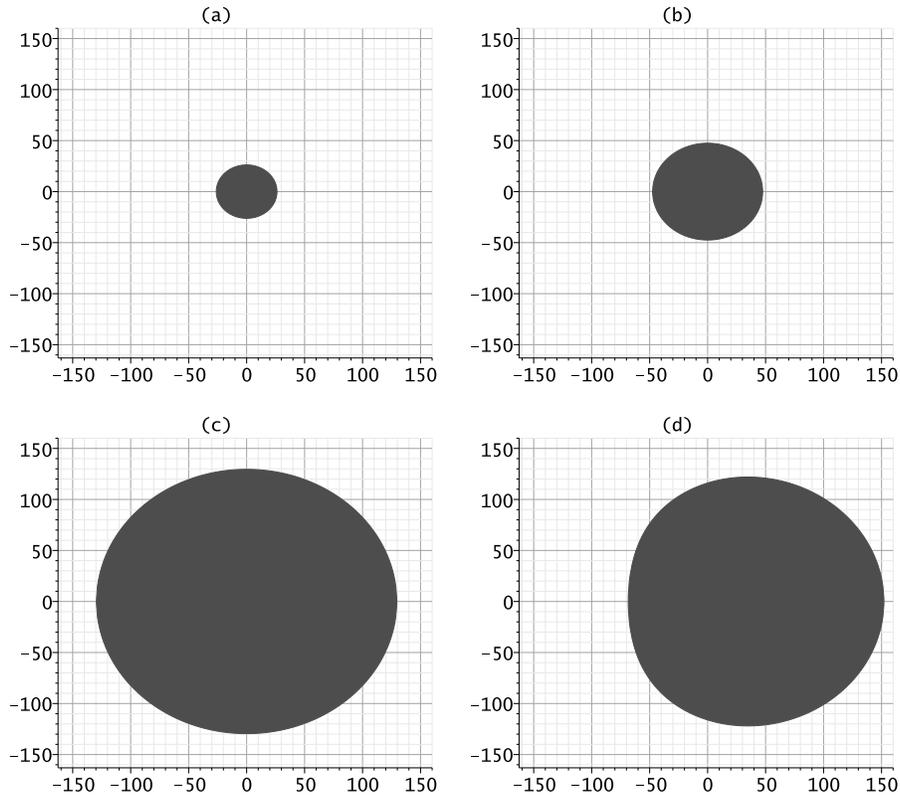}
\caption{The expected size of the shadow (in microarcseconds) of the photon sphere of Sgr A* for different values of the MOG $\alpha$ parameter: a) $\alpha=0$, b) $\alpha=1$, c) $\alpha=5$, assuming no rotation. For comparison, d) shows near extremal rotation at $\alpha=5$ and $a=0.99GM$.\label{fig:SgrA*}}
\end{figure}

In the non-rotating case $a=0$, the photon sphere radius is given by
\begin{align}
r_\gamma=\frac{1}{2}(1+\alpha)G_NM\left(3+\sqrt{\frac{9+\alpha}{1+\alpha}}\right)=\frac{1}{2}(1+\alpha)G_NMX,
\end{align}
where, following \cite{Wang2019} we introduced the shorthand $X=3+\sqrt{(9+\alpha)/(1+\alpha)}$. Furthermore, at $r=r_\gamma$, we have
\begin{align}
\Delta_\gamma=r_\gamma^2-r_sr_\gamma+r_Q^2=\frac{1}{4}(1+\alpha)^2G_N^2M^2X^2-(1+\alpha)^2G_N^2M^2X+\alpha(1+\alpha)G_N^2M^2.
\end{align}
The shadow radius, in turn, is determined by
\begin{align}
r_{\rm shadow} = \frac{r_\gamma^2}{\sqrt{\Delta_\gamma}}=
\frac{(1+\alpha)^2X^2}{2\sqrt{(1+\alpha)^2X^2-4(1+\alpha)^2X+4\alpha(1+\alpha)}}G_NM.\label{eq:rsh}
\end{align}

\begin{figure}
\includegraphics{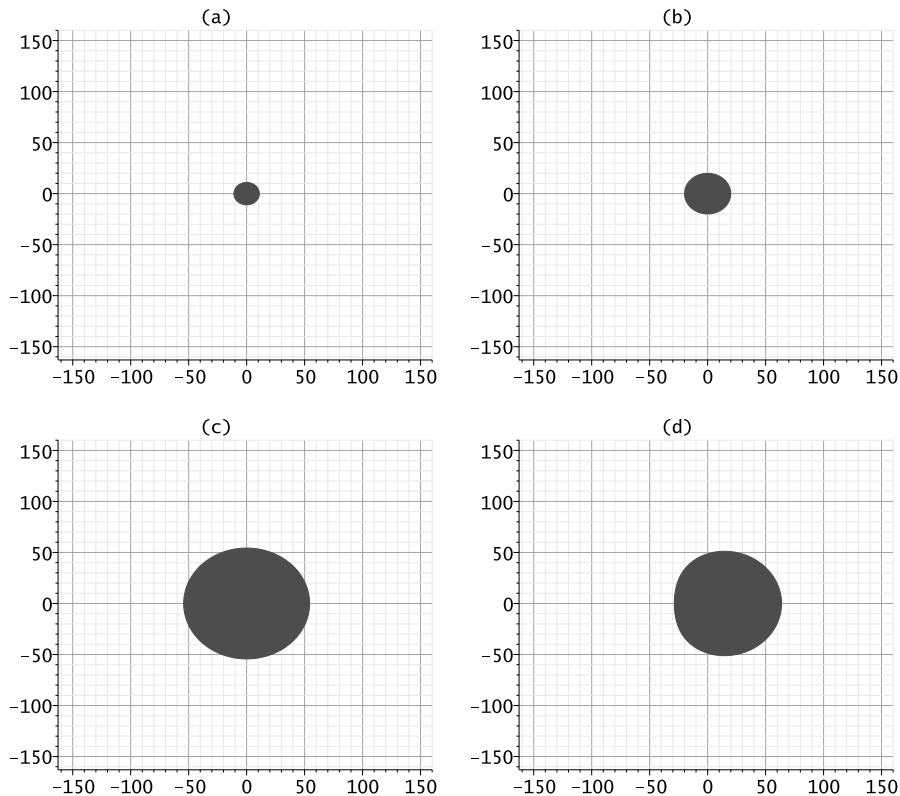}
\caption{The expected size of the shadow of the photon sphere of the supermassive black hole in M87 for different values of the MOG $\alpha$ parameter: a) $\alpha=0$, b) $\alpha=1$, c) $\alpha=5$, assuming no rotation ($a=0$), and d) $\alpha=5$ and $a=0.99GM$.\label{fig:M87}}
\end{figure}
\end{widetext}

In Fig.~\ref{fig:SgrA*}, we show the apparent size of the shadow of Sgr A* for different values of $\alpha$. For this, we adopt the value of $M_{\rm Sgr A*}=4\times 10^6~M_\odot$ for the Newtonian mass of Sgr A*, and assume a distance of 7.86~kpc from the Earth \cite{Boehle2016}. For comparison, we also include a near extremal rotating case, using the formulation presented in \cite{Moffat2015b} (not reproduced in the present paper; see also \cite{Guo2018}.)

As this figure demonstrates, and as can also be seen from Eq.~(\ref{eq:rsh}), the shadow radius for $\alpha>1$ scales approximately proportionately to $\sim (1+\alpha)$. In other words, whereas Keplerian dynamics in the vicinity of the black hole are governed by the Newtonian mass in the expression $G_NM$, the size of the black hole shadow is proportional to the ADM mass, determined by $\sim(1+\alpha)G_NM$. This offers a method to determine observationally the value of $\alpha$, characteristic to a specific black hole.

In addition to Sgr A*, we also computed the anticipated shadow size for the supermassive black hole in M87. Determinations of the Newtonian mass of the M87* black hole are far less reliable than those of Sgr A*, as no individual stellar orbits can be resolved. Widely adopted values around $7\times 10^9~M_\odot$ \cite{Gebhardt2011,Oldham2016} are dynamical mass estimates that take into account the dynamics of the entire galaxy, modeled using assumptions about its mass-to-light ratio and, more importantly, the presumed size and mass distribution of its dark matter halo. This is clearly in conflict with the concept of modified gravity without dark matter, thus we rely instead on mass estimates that do not depend on any assumption related to dark matter and other properties of the galaxy, but are based instead on the dynamics of gas in the central region of M87: $M_{\rm M87*}=(3.5^{+0.9}_{-0.7})\times 10^9~M_\odot$ \cite{Walsh2013}. Accordingly, to compute the size of the M87* shadow, we adopted this value as the M87* black hole mass, located at a distance of $(16.4\pm 0.5)$~Mpc \cite{Bird2010}. The results of this calculation are shown in Fig.~\ref{fig:M87}.

The value of $\alpha$ for these supermassive black holes is not known. Assuming that the semi-empirical derivation developed in \cite{Moffat2007e} remains at least approximately valid, we anticipate $\alpha<1$ for Sgr A* and $1<\alpha$ for M87*.

\begin{table}
\caption{Expected values of the angular diameters of Sgr A* ($\delta_{\rm Sgr A*}$) and M87* ($\delta_{\rm M87*}$, assuming the gas dynamics mass estimate of $M_{\rm M87*}\sim 3.5\times 10^9~M_\odot$) as functions of the MOG parameter $\alpha$. NB: $\alpha=0$ corresponds to general relativity.\label{tb:R}}
\begin{tabular}{c|c|c}
~\hskip 1.5em$\alpha$\hskip 1.5em~&~\hskip 1.5em$\delta_{\rm Sgr A*}$\hskip 1.5em~&~\hskip 1.5em$\delta_{\rm M87*}$\hskip 1.5em~\\
~&$(\mu{\rm as})$&$(\mu{\rm as})$\\\hline\hline
\phantom{1}0&48.4&22.0\\
\phantom{1}1&87.8&39.8\\
\phantom{1}2&126.1&57.2\\
\phantom{1}3&164.1&74.4\\
\phantom{1}4&201.9&91.6\\
\phantom{1}5&239.5&108.6\\
\hline
\end{tabular}
\end{table}

Values for the anticipated angular diameters of the Sgr A* and M87* black holes for different values of $\alpha$ are shown in Table~\ref{tb:R}.

The inferred size of the shadow of the M87* black hole has just been revealed by the Event Horizon Telescope project \cite{2019EHT1,2019EHT6}. The angular resolution of this VLBI observational campaign was sufficient to resolve the shadow of the M87* supermassive black hole, estimating its angular diameter at $42\pm 3$ microarcseconds, corresponding to $\alpha=1.13^{+0.30}_{-0.24}$ using the mass estimate obtained from the dynamical gas model \cite{Oldham2016,Walsh2013}.

\section{Conclusions}

Thus we conclude that the MOG theory, with $\alpha\sim 1.13$, is consistent with the observed angular diameter of the M87* black hole's photon sphere shadow and the M87* mass obtained from gas dynamics. Indeed, as there is no dark matter in the MOG theory and, therefore, mass estimates for the M87* black hole from stellar dynamics may be lower, the theory may help resolve the apparent conflict between mass estimates derived from gas vs. stellar dynamics.

In addition to the gas dynamics based mass estimate, two additional mass estimates for M87* recently appeared in the literature. In \cite{Titarchuk2019}, correlation between the spectral index and mass accretion is used to derive the very low mass estimate of only $M_{\rm M87*}=5.6\times 10^7~M_\odot$, which would correspond to an excessively large value of $\alpha=153.8$ to produce a shadow with an angular diameter of $42~\mu{\rm as}$ in the MOG theory. In contrast, the authors of \cite{Nokhrina2019} used the shape of the M87* jet boundary to derive $5.2\times 10^9~M_\odot\lesssim M_{\rm M87*}\lesssim 7.7\times 10^9~M_\odot$, the actual value dependent on a choice of a magnetization parameter. These estimates correspond to $0\lesssim\alpha\lesssim 0.35$.

~\par

\begin{acknowledgments}
We thank Martin Green for helpful discussions. This research was supported in part by the Perimeter Institute for Theoretical Physics. Research at the Perimeter Institute is supported by the Government of Canada through the Department of Innovation, Science and Economic Development Canada and by the Province of Ontario through the Ministry of Research, Innovation and Science.
\end{acknowledgments}

\bibliography{refs}
\bibliographystyle{apsrev}

\end{document}